# A Survey of Routing Techniques for Mobile Collaborative Virtual Environment Applications

Anis Zarrad[1]

[1] Computer Science and Information Systems Department, Prince Sultan University
Riyadh, Saudi Arabia
*azarrad@psu.edu.sa*

**Abstract**
Due to the rapidly increasing usage of personal mobile devices and the need of executing CVE applications in environments that have no previous network infrastructure, **M**obile **C**ollaborative **V**irtual **E**nvironments (MCVEs) systems will become ubiquitous in the future. In such systems, we enable users to act and interact in three dimensional world shared through their mobile devices in an ad-hoc network (MANET). A handful of interesting MCVE applications have been developed in a variety of domains, ranging from multiplayer games to virtual cities, virtual shopping malls, and various training simulations (i.e. military, emergency preparedness, Education, Medicine, etc.). The designers of such applications rely solely on network overhead, latency, limited bandwidth, and mobile device limitation. In this survey we present a number of ways to classify and analysis routing techniques that have been applied in Mobile Collaborative Environment Applications.

## 1. Introduction

In recent years, the usage of mobile technologies in collaborative virtual environment has received increasing attention throughout the computing community. In Mobile Collaborative Virtual Environment MCVE users can interact with shared virtual environment using their mobile devices, which are interconnected through an ad-hoc network (MANET). A graphical body called an avatar represents each user. Through the avatar concept, players are able to see and hear each other. Many potential MCVE applications have been used, such as online games, virtual shopping, training simulations, and emergency preparedness. Figure 1 shows an example of a MCVE application - Emergency preparedness Fire scenario.

Mobile access to the above-mentioned environments can be quite appealing since users must have continuous access to services. Network overhead, scalability, and data persistency place a major impact on the performance of any MCVE application. Persistency is defined as retaining the world state from one play session to another even when a player is offline without losing any relative data.

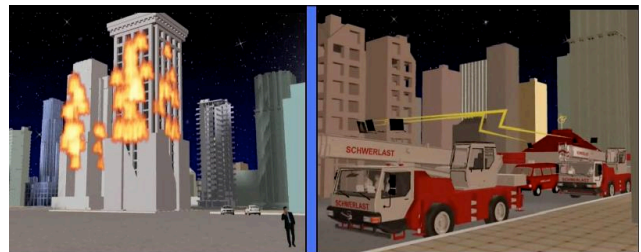

Fig .1: Mobile Collaborative Virtual environment overview

Most Collaborative virtual environment technologies works well on wired networks and desktop settings where network resources are so plentiful, therefore they are not adequate enough for addressing mobile collaborative virtual environments. The main objective in MCVE application is that it should be possible to interact with the environment from anywhere by everyone even when a network infrastructure is not available due to a disaster or military use, and to do so CVE requires to work over MANETs.

In MCVE applications, one of the main research topics for such environments is how to efficiently transmit messages in order to minimized network overhead, and network delay. These applications must send information over the MANETs in real time in order to sustain the felling of immersion while user interacts with the environment. During such transmission the information is subject to network loss, and network latency, which may cause several problems in these types of applications. Messages must be delivered with an acceptable network delay [5], and [6], with higher latencies, the user starts losing the feeling of immersion, decreasing his/her rate of interaction, and eventually causing him/her to lose interest to the application. Therefore, routing techniques in MCVE system gradually becomes an important focus part of designing such system, because a weakness in routing might lead unpleasant results.

In this paper we begin by giving a general overview about Collaborative virtual environment and its applications. The rest of the paper is organized as follows. First we summarize the evaluation approach we use in our review of MCVE applications, second we present and analyze some of the major MCVE systems to date and their classification, and finally in section four we offer conclusions.

## 2. Backgrounds and Related Works

The main focus in this survey is mobile collaborative virtual environment; however, since we believe there is a need to highlights the most influence works related to virtual environment systems occurred in the past 40 years in order to give credits to founder researches. The topic of virtual environment starts in the late 1950's [1] pioneering the concern to change the way people interacted with computers and makes possible virtual reality. Flight simulator [7] was introduced as one of the most significant antecedents of virtual reality. After that a number of studies have led the implication of human–computer interaction applications.

In this section we provide an overview of several desktop CVE systems ranging from military training combat (land, earth, sea) [8] to commercial application [10], [13] to multiplayer games [12], [11], and virtual shopping mall [9], etc. We classify our studied CVEs applications into two main categories: *peer-to-peer* and *client server* models. CVEs designers can choose one model, or they can 'mix' both models, using peer-to-peer for some aspects of their design, and client-server for others. Figure 2 depicts the main communication architecture solutions used in the literature to develop CVEs applications.

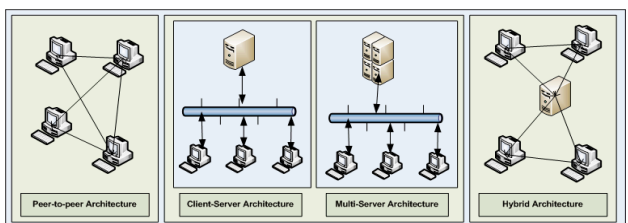
Fig. 2 CVE Networked Topology

Most of the studied CVEs systems were developed either for academic or commercial applications. Here we discuss some of them.

### 2.1 Client-Server-based CVE

In this section we present the main CVE systems that fall in the client-server and multi-server categories. Each avatar has to send an update message to the server, no matter how many users are participating in the CVE, and receive an update message from other nodes through the server. The major users of client server model are commercials multiplayer games; EverQuest [10], and UltimaOnline [11] are such games worthy of mention. Usually, game designers choose to exhaust less effort on network performance in favor of enhancing graphics representation and sounds. Many other works have been proposed in the CVE literature [15], [16], [17], and [14] for academic and commercial use.

BrickeNet [14] provides a filtering technique to minimize the number of messages to be handled by each user. Compared to other existing architectures, BrickeNet introduces an interesting strategy of controlling shared objects. Instead of sharing the virtual environment, each node manages a local copy of the VE, and selects a set of objects to be shared with other users. This strategy offers the user full privacy, but only at the expense of collaboration activities, since users cannot access all the VE resources. NetEffect [64] was designed to simplify the development of network-based virtual worlds. A server master guarantees the load balancing among the set of available servers by migrating clients from one overloaded server to another one with less workload.

In Ring system [17], authors exploit a Potentially Visible Set (PVS) data structure to handle the scalability requirement. Unlike NetEffect, Ring uses a static virtual environment partition, where each user is represented by an entity rendered on every other computer participating in the network. Based on its current position, the user enjoys the leisure of selecting statistically which server to connect to. The key feature of the Ring system is that the server-based visibility algorithms compute potential visual interactions between users in order to reduce the number of messages required to maintain consistency. Peer-to-peer architecture has been developed as an alternative solution to the Client-Server; next section describes some peer-to-peer CVE applications.

### 2.2 Peer-to-peer Network based CVEs

In peer-to-peer systems, each avatar directly communicates its update information to all other participants on the network. In such a model, there is no central server and no single point of failure, since all nodes play client and server roles. A number of well-known applications, namely DIVE [21], MiMaze [25], NPSNET [26], SPLINE [27], SCORE [24], and VON [23] are extensively described; other, less common applications like MASSIVE [22], Federated peer-to-peer [20], and SimMud [19].

Federated peer-to-peer [20] uses hybrid architecture, where nodes are organized into various groups, and managed by a *multicast reflectors* scheme. SimMud [19] proposes a solution based on DHT Pastry [28] structured network to support massive online multiplayer games. The VE is divided into regions, with each region managed by a super node, and playing the root role for the multicast region tree. Any user action is received by the root and delivered to all multicast members. Network latency is affected, since communication between nodes operates under DHT [18].

MASSIVE [22] (**M**odel, **A**rchitecture, and **S**ystem for **S**patial **I**nteraction in **V**irtual **E**nvironments) proposes a technique to limit the number of connections as the number of users increases, in order to reduce network traffic. Each object is associated with an *Aura* that defines the area in the VE in which the object can publish. Objects can communicate only when their *Auras* overlap. MASSIVE was developed to handle mainly scalability and network heterogeneity. To the best of our knowledge, **NPSNET** [26] (**N**aval **P**ostgraduate **S**chool NPS) and **SPLINE** [27] (*Scalable Platform for Large Interactive Networked Environment),* represent the most well well-known peer-to-peer CVE prototypes.

**NPSNET** [26] was developed in 1990 for large-scale military simulations using the Model-View-Controller (MVC) pattern to offer reusability and simplicity. NPSNET uses SIMNET and DIS [26] as networking technique solutions to interoperate with other simulation systems. The virtual environment is divided into a well-defined hexagonal cells structure, with each cell having its own multicast group so as to save network bandwidth. NPSNET has been designed with special focus on military simulation application. An improvement version of NPSNET called **DIVE** [21] (*Distributed Interactive Virtual Environment*) was developed at the Swedish Institute of Computer Science. DIVE uses RTP (Real-time Transport Protocol) for stream data and SRM (Saleable Reliable Multicast) for non-stream-based data communication. The network traffic is reduced by means of the SRM protocol, since there is no need to send keep-alive message. The SRM protocol is able to detect missing data packets on reception. We classify DIVE under the hybrid architecture category since there is a central node called *DIVESERVER*, which is needed to supply the initial connection; however, Frécon *et al* classified it as a peer-to-peer system.

**SPLINE** [27] (*Scalable Platform for Large Interactive Networked Environment),* was developed in order to provide a solution that facilitates interoperability between system parts. SPLINE enables a hybrid topology based on the client-server and multicast approach in order to cope with low bandwidth networks. In the proposed architecture, users do not communicate with each other, instead they only communicate with the world model. In contrast to the regular partitioning scheme used for NPSNET [26], SPLINE [27] divides the virtual world into arbitrary shapes called *LOCALES* where each *LOCALE* has its own multicast. Users are permitted to be "present" in more than one *LOCALE* through the use of *spObserver* objects. *LOCALES* are considered to be the key feature for improving scalability in SPLINE scalability. There are many other works that can be stated such as **SCORE** [24], **MiMaze** [25], and **VON** [23].

## 3. Distinctive Classifications of MCVEs

Due to the manner in which the network overlay is constructed, ad-hoc routing may not be efficiency employed in terms of discovery service and number of underlay hops. Therefore we have created a novel classification for MCVE applications. Our classification space provides methods for comparing MCVE systems according to three key characteristics: *Network overlay topology*, *Routing techniques* and *Type of experimental results produced*.

In this survey our stated goal is to organize various routing technologies applicable to build MCVE applications into several categories. The following classification offers a convenient way to group and compare methodologies, and challenging issues in MCVE applications. It is not the case that one category is necessarily better than another: advantages and disadvantages exist in all categories. As a first set, network overlay may be separated into two main classes: **Structured** and **Unstructured**. A second set related to MCVE evaluation is the routing technologies used to build such applications. There have been several routing protocols proposed for MANETs in the literature. In this paper we classify them based on the criteria **Reactive**, **Proactive** and **Hybrid**. In Reactive scheme also called on-demand the route is determined when needed, whereas in proactive scheme also called table driven the route is determined in advance, so that the route is present whenever is needed. Hybrid scheme combines the advantage of reactive and proactive.

3.1 Structured Overlay Network

In structured overlay network, specific nodes strictly control the topology. The lookup service is based on a distributed hash-table (DHT) [18] to efficiently locate the data in the network. There are many structured overlay proposed in the literature, however in this survey we limit our discussion to the most well known ones used in MANETs: Pastry [28] and Chord [29]. Adapting structured overlay for MANETs has attracted more researches in the recent years; however the major application area of such approaches is about studying the performance behavior of protocol itself [64, and 66], applicability to a real scenario such as disaster, battlefield etc is not practical. The fundamental concept is to improve the routing performance by exploring DHT in MANETs. What follows is the discussion of some relevant applications and their classification based on the routing techniques.

3.1.1 Reactive Routing Scheme

We describe in this section some approaches that uses reactive routing scheme.
In [30] authors studies the performance of Pastry like algorithm over MANETs using hop-by-hop routing AODV [31]. Simulation results has shown a poor network performances due to heavy network overhead need it to maintain large numbers of connections between nodes. An alternative solution may be proposed to overcome this weakness by introducing a cross layer approach.

Hu *et al* [35] propose an evaluation of a scalable solution over MANETs. The idea is based on a combination of the DSR [32] and Pastry [28] to create Dynamic per-to-peer source routing called **DPSR**. An ID is assigned to each node using a specific hash. Messages are routed based on the hash address destination. This strategy enable fast look up resolution in spite of highly dynamic nodes.

Pusha *et al* [64] provide an efficient way to construct a mobile file sharing application using Pastry [28] and DSR [32] in an integrated approach. With this integration, the routing structure in DHT is integrated with the routing cache of DSR into one structure to optimize the routing performance and to learn about paths to close nodes. Experimental simulations have delivered promising results, which demonstrate the feasibility of this approach.

3.1.2 Proactive Routing Scheme

This section describes some relevant works that uses Proactive scheme as ad-hoc routing protocols.

In Cross- ROAD [65] Delmastro present an extended version of OLSR with structured overlay functionality by hashing the IP addresses in the routing table. In this work authors compared CrossROAD's performance with the Pastry overlaid on OLSR. Simulations results show that the proposed cross-layer modification eliminate the network overhead required for the overlay maintenance.

Cramer and Fuhrmann [66] have evaluated the behaviour of Chord lookup protocol in MANET using the network simulator GloMoSim [67]. Various routing protocols Ad hoc On-Demand Distance Vector (AODV) routing [31], Dynamic Source Routing (DSR) [32], and Optimized Link State Routing (OLSR) [31] are used in the simulation scenarios. Experimental results show poor performance and unpredictable behaviors when we increase the network size.

MobiGrid [37] address the problem of information sharing using an interplay scheme between structured peer-to-peer system and MANET. This architecture is unique in that it allows nodes to negotiate the key space in order to build a sophisticated binary search tree.

3.1.3 Discussion

There are some other studies about using structured overlay networks over MANETs using different class of routing, like geographical routing as proposed in [68]. The protocol design is based on Geographical Hierarchical Index (GHI) over CAN.

Efforts are often made for mobile file sharing applications, which represent a target for MCVEs. For the purpose of determining their performance, the structured overlay network is undesirable for mobile ad-hoc networks because node mobility will hamper their routing efficiency, especially when dealing with application that require a user to participate in a battlefield situation or emergency preparedness scenario after an earthquake. The most common problem of the structured network is that it does not address the area of interest [23]. Also DHT [18] may suffer more overheads in the maintenance process; an unstructured network may therefore perform better.

3.2 Unstructured Overlay Network

Unlike structured networks, the topology in unstructured networks is created arbitrarily. The unstructured peer-to-peer model comprises three generations. The first (Centralized) is a mixture of client-server and peer-to-peer networks; the second (Decentralized), is a pure peer-to-peer network where nodes can play both roles (*client or/and server*); and the third, a hybrid approach of the first and second generations. Table 1**Error! Reference source not found.** provides an overview classification of the studied unstructured peer-to-peer network architectures. Many of them are widely used in this area, namely Napster [38], JXTA [48], and Gnutella [40]. Others are less common like Kazaa [41], and Morpheus [42].

Table 1: Unstructured Peer-to-peer System Classification

|  | Centralized | Decentralized | Hybrid |
|---|---|---|---|
| Kazaa |  |  | ✔ |
| JXTA |  |  | ✔ |
| Napster | ✔ |  |  |
| Gnutella |  | ✔ |  |
| Morpheus |  |  | ✔ |

Most attempts to combine peer-to-peer networks and MANETs using a cross layer approaches. At the present time, researchers focus their efforts on mobile file sharing applications as mobile collaborative applications. What follows is discussion of some approaches based on the classification presented in Table 1.

3.2.1 Centralized Network Overlay

In centralized architecture, a server is dedicated to maintain a peer's index and resources. The server processes all resource queries, and returns a matching list to the requesting node.

Napster on the road [3] is perhaps the most well know mobile file sharing application. It was based on a centralized architecture where a central server maintains the database of song files. Mobile customers can browse Napster's music catalogue and sample any one of the tracks available via a free preview of up to 30 seconds. Mobile Napster is a premium music service in Europe reflecting a very dynamic market for digital and mobile music.

3.2.2 Decentralized Network Overlay

No dedicated servers are required in a decentralized approach. A broadcast transmission is required for both, peer discovery and query requests. Gnutella [40] is taken as leader of decentralized approaches. Follows are some approaches:

Choi et al [43] outline an enhanced Gnutella protocol for ad-hoc networks that addresses the PONG message flow problem. A peer-to-peer metric value is used to select the Ultrapeer node candidate. The proposed system provides better performance than Gnutella in terms of query hits and network overhead.

In [43], and [44] authors outline a cross layer technique to enhance the Gnutella network in mobile environment. Boukerch et al [50] illustrate possible solution to deploy Gnutella network over MANETs to support MCVE applications. The developed protocol is based on cross-layer approach between the application and the network layer to adopt the user interest in virtual environment.

Simulation results has shown that the system provide a better performance when implemented over HSR compared to AODV, DSR and ZRP.

Tang et al [69] propose an integrated approach for peer-to-peer file sharing on multi-hop wireless networks. The approach relies on implementing FASTTRACK adopted by KaZaA over Ad Hoc On-Demand Distance Vector protocol (AODV) protocol [31]. Nodes use FASTTRACK routing to receive the node ID storing the requested file. Then, AODV is charged to find the best route to that node. Performance analysis shows a reduced average delay perceived by each file requester, and reduced network overhead.

Seneviratne et al [45] addresses the problem of mobile file sharing by introducing the mobile agent. The architecture uses mobile agents to participate in Gnutella network on behalf of mobile devices, in order to reduce the volume of communications messages and the power consumption of the mobile devices.

A pure peer-to-peer (i.e. without super-peers) network over MANET [46] was used by the PROEM [47] project. The proposed approach may be used for diverse mobile application like file sharing, and instant messaging. PROEM messages are based on TCP, UPD and HTTP, and it employs XML technology for their messages representation. The main limitation of the proposed approach is poor performance when the network size is increased.

3.2.2 Hybrid Network Overlay

Hybrid approach is the advantage combination of the centralized and decentralized schemes. JXTA [39] is a well-known example of a hybrid approach where only selected nodes, called super-peers, are used for resource discovery and query process.

Traditional JXTA [39] cannot be a feasible solution for mobile applications due to message flooding, network heterogeneity, and wireless link reliability problems. As a result, the JXTA community developed a light version in order to support the mobility requirement. Two different platforms were designed: JXME Proxied and JXME Proxyless. Proxyless is still under construction, and thus, a number of design and implementation issues need revision. The JXME Proxied platform is based on a hybrid peer-to-peer mode, it relies on central entities in order to play a proxy role between nodes belonging to the same JXME virtual network, and all mobile devices must be in proximity to the server (JXTA relay) so that they can communicate through this server. This approach, however, considers only cellular network and does not aim to match the virtual network to its physical counterpart. Another problem with JXME is interoperability; JXME is made exclusively for J2ME (Java 2 Micro Edition).

PnPAP [49] (Plug and Play Application Platform) is a hybrid system solution through which applications can access different types of networks over various protocols (SIP, JXME [39], and Jabber); PnPAP offers free opportunity and simplicity since it allows users to employ many network technologies simultaneously in the network layer.

3.2.4 Reactive Routing Scheme

In this section we describe some approaches that uses proactive ad-hoc routing protocols.
ORION system [51] explores the concept of cross-layer over MANET as a scheme combining the discovery process routing at the network layer and the query process at the application layer. The basic of ORION is based on the AODV routing protocol. Performance results has shown unnecessary network overhead due to application layer routing.

MPP [48] (Mobile Peer-to-peer Protocol) is another example of cross layer approach, and is based on DSR [32] ad-hoc routing protocol. This protocol establishes a communication channel between application layer and network layer to deploy peer-to-peer system effectively in mobile environment. Each node announces itself to the routing protocol, and a search request is broadcasted to all nodes. MPP relies on flooding for query lookup. It is clear to see that MPP does not scale when the network size and the network query rates increases.

3.2.5 Proactive Routing Scheme

In this section we present some significant approaches used in this area that uses proactive ad-hoc routing protocols.

Guettier *et al* [70] propose an approach in an ad-hoc network to support multiple collaborative services. The protocol relies on OLSR [31] routing to address mobile tactical communications, and to provide proactively a route always available between any two nodes. The key concept of this protocol is the use of specific parameter MPR-COVERAGE to deal with the unnecessary network overhead. Received message is checked based on the parameters to decide about it's state (Drop or Forward). Simulation results reveal better performance in term of network latency, and dropped network packets.

Ready *et al* [63] describes an Efficient DSDV Routing Protocol for Wireless Mobile Ad Hoc Networks called *Eff-DSDV* to overcome the problem of stale routes, in regular DSDV. Each entry in the routing table has an additional entry for route update time. This update time is used to verify the route state whenever is possible. Simulation results have shown better performance compared to regular DSDV when we consider packet-delivery ratio, end-end delay, dropped packets, and routing overhead metrics. However, managing the update date requires extra efforts.

3.2.6 Hybrid Routing Scheme

This section describes some approaches used in mobile environment over hybrid ad-hoc routing protocols.

CHAMELEON [62] is a hybrid protocol developed for multimedia communication in emergency case. It was designed to adapt its routing behavior according to the size of MANET. A combination between the reactive Ad Hoc on-Demand Distance Vector Routing (AODV) and proactive Optimized Link State Routing (OLSR) protocol is used. Simulations results show that CHAMELEON has an overall improved delay and jitter performance over both AODV [31] and OLSR [31] routing protocols.

Christopher *et al* [61] present an extended version of ZRP [36] protocol in order to improve the service discovery in MANETs. The approach implements a service discovery in the routing layer by piggybacking the service information into routing message, in doing so we decrease communication network overhead and save battery power. Based on their simulation results, extended ZRP provides better performance compared to ZRP.

Halgo *et al* [60] describe an emulation environment tool called MASSIVE (MAnet Server Suite Incorporating Virtual Environments) for mobile ad-hoc networks to perform a system analysis. The main focus lies on; analyze the interaction between of the ad-hoc network and the applications itself. As an example authors incorporate an enhanced routing protocol TORA with an E-Learning in ad hoc network application [58].

3.2.7 One Hop Routing Scheme Face-to-Face
This is a very commune technique used by researches to design face-to-face collaboration system. In this technique nodes should be close enough in order to communicate and collaborate with each other. What follows is a description of some of these schemes in order to improve the routing performance.

Alf *et al*. [53] designed a framework to support Computer-Supported Cooperative Work (CSCW) on mobile phones using J2ME and Personnel Area Network (PAN). The developed scheme is not suitable because CVE systems are expected to enable user interaction, and collaboration with the environment and other participants regardless of their proximity. CSCW is still under progress; simulation scenarios are very limited (only three users).

iCoulds [52] is a simple peer-to-peer information-sharing system for mobile networks. It was developed mainly for J2ME (Java 2 Micro Edition). The communication layer in iClouds architecture is based on a simple one-hop exchange message. Proximity limitation and *information sprinklers* are the limitations of this project.

Anhinga *et al* [59], propose an approach that does not make use of an ad hoc routing protocol. Instead it technology is based on proximity and community systems to support face-to-face collaboration in mobile ad hoc networks. The system design relies on lightweight Java Version, Jini Network, and tuple spaces. The proposed solution runs on groups of proximal mobile devices characterized by many-to many communication patterns.

3.2.8 Other Unclassified Approaches

The presence of high variability of ad-hoc routing protocols in the computing area, however, does not always allow researchers to reach their goals due to complexity of the task and distinctive characteristics of system. This could be addressed by developing new methodologies and protocols, so system performance can be improved. What follows is a discussion of some of these schemes.

Bejan *et al* [57] define a mobile platform for online games. A separate session node manages each game. The discovery process relies on a Resolver Service to send queries and process search services. Such service offers better management solution for mobile games. The Resolver Service is an interesting alternative, but introduces high network overhead, since each device must be queried about whether or not it will host the service.

A promising social application called MobiTrip [9] was developed to allow many users to express their opinion in the environment. User options are made available when user devices connect on the fly, or when users approach connection hotspots. Bluetooth technology is used in order to form a social space of nearby devices. A typical example of such an application is the shopping mall, where users can express their opinion about new products.

Michael *et al* [58] propose an information services supporting E-Learning application over MANETs. The emphasis of this work is to deal with the flooding problem in wireless network. Therefore, a new service discovery mechanism is developed using a combinations idea of content based routing protocol [2] and cluster head proposal [4] to limit the network overhead.

3.2.9 Discussion

Apart the approaches mentioned above there are many other relevant mobile applications, such as PDA access to video stream, remote processing, remote controllers [54], inter-vehicle communication, and vehicle-to-road communication [56] that have been reported in the literature.
All in all, while most of the existing proposed mobile systems in the literature are greatly beneficial (and applicable) in the research academic and commercial sectors, they falter suffer from being limited to specific tasks, and their architectures are typically tightly coupled to the characteristics of their networks, and any extension or modification to adopt a new class of application is almost impossible.

In some discussed solution authors state different opinions about the performance when combine peer-to-peer and MANETs. Barbosa *et al* [55] describe an interesting evaluation of ad-hoc routing protocols under Gnutella peer-to-peer network. Their simulation concludes that none of the ad-hoc routing protocols performed well under all circumstances in Gnutella network.

## 4. Conclusions

In this survey we have presented various criteria for classifications of Mobile Collaborative Virtual Environment (MCVE) applications. There are still many challenges facing MCVEs especially when developing real-time application. The increasing popularity of peer-to-peer system and MANETs gives more attention to researches to combine both systems in order to built efficient Mobile collaborative Virtual Environment applications. Peer-to-peer networks and MANETs [24] do share some similarities, such as decentralization, self-reorganization, and self-healing; the idea of combining both networks is therefore a natural decision. However, this combination poses great challenges because these networks operate on different layers (peer-to-peer at the application layer and MANET at the network layer).

As far as the author is concerned, there has been no report on CVE applications in mobile peer-to-peer networks. Many works classify mobile file sharing application as a MCVE applications. It is important to mention that there is a notable difference between a file-sharing application and a CVE application. In the latter, the content search and the computing are much more complex, as the user interest may vary over the time. Also network performance has an important impact on collaborative application compared to file sharing application.

In MCVE application many complicated data such as 3D scene background, and Avatars data need to be exchanged which make the system more complex to manage. Also data persistency has been a major issue ever since user can disconnect suddenly. It is the central issue of any MCVE applications.

**Acknowledgments**

The author is thankful to Dr. Ezzedine Boukerch, and also to Prince Salman Research and Translation Center for providing the necessary support for the preparation of this work.

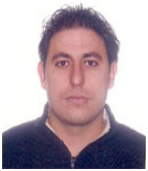
**Anis Zarrad:** is an Assistant Professor at the Computer Science and Information Systems Department in Prince Sultan University, Riyadh, Saudi Arabia. Dr. Anis research interest are: Software engineering, peer-to-peer networks, and Mobile Collaborative Virtual environment to model emergency properness scenarios.  Context aware computing and wireless sensor networks are integrated to control and visualize physical environments subjects to emergency conditions. Dr. Anis has published several research papers in conferences and international journals.

Dr. Anis Zarrad serves as an editor in several international journals.